\documentclass[a4paper,11pt]{article}
\usepackage{jcappub} 
\usepackage{indentfirst}
\usepackage{epsf,amsmath,graphicx,amssymb,url,hyperref,dcolumn}
\usepackage{mathtools}
\usepackage{color}
\usepackage{longtable,tabu}
\usepackage[normalem]{ulem}
\usepackage{adjustbox}
\usepackage{array}
\usepackage{xcolor}

\title{\boldmath Constraints on Axion-Photon Mixing from Fast Radio Burst Dispersion Measures}
\author[a]{Gunalan Muthusami}
\author[a,1]{and Gopal Kashyap\note{Corresponding author.}}
\emailAdd{gunalan.2024@vitstudent.ac.in}
\emailAdd{gopal.hari@vit.ac.in}

\affiliation[a]{ Department of Physics, School of Advanced Sciences,
Vellore Institute of Technology, Vellore, Tamil Nadu 632014, India }

\abstract{
Fast radio bursts (FRBs) offer a powerful probe of the ionized Universe through their dispersion measures (DM). While a significant fraction of the DM arises from the intergalactic medium (IGM), the contributions from the host galaxy and the immediate environment of the source remain uncertain, and the physical origin of FRBs is still under active investigation. In this work, we investigated the possibility that FRBs originate from high-magnetic-field neutron stars (NS), whose magnetospheres can facilitate axion-photon mixing. Such mixing can modify photon propagation and induce an effective contribution to the observed dispersion.
Using a sample of localized FRBs with measured redshifts, we perform a Bayesian Markov Chain Monte Carlo (MCMC) analysis to constrain the axion mass $m_a$ and axion-photon coupling $g_{a\gamma\gamma}$. Within a parametric cosmological framework, we obtain
$m_a = 1.16^{+4.40}_{-1.08}\,\mu{\rm eV}$ and 
$g_{a\gamma\gamma} = (1.76^{+6.69}_{-1.64})\times10^{-16}\,{\rm GeV}^{-1}$, 
together with a physically consistent intergalactic baryon fraction 
$f_{\rm IGM} = 0.837^{+0.053}_{-0.056}$. 
We further tested the robustness of our bounds against cosmological modeling assumptions by employing a non-parametric Gaussian Process reconstruction (GPR) of the DM-$z$ relation, which gives statistically consistent results.
}
\keywords{fast radio bursts, dispersion measures, redshifts, intergalactic medium, axions, neutron star, magnetospheres}

\begin{document}
\maketitle
\section{Introduction} \label{secmagnetospheres:Intro}
Fast radio bursts (FRBs) are extremely luminous, millisecond-duration radio transients among all other known astrophysical radio signals  \cite{KATZ20181,lorimer2024discovery,RevModPhys.95.035005, https://doi.org/10.1002/asna.20250024}. The large dispersion measures (DM) observed in FRB signals suggest that their sources are predominantly extragalactic \cite{kulkarni2020dispersionmeasureconfusionconstants,PhysRevD.100.083533}. Despite significant observational progress, the physical origin of FRBs remains an open problem in astrophysics. At the same time, the excess DM measured in FRBs provides a useful probe of the distribution of ionized matter content distributed along the line of sight \cite{macquart2020census}. The observed DM encodes the column density of free electrons integrated over the propagation path, including contributions from the interstellar medium (ISM) of the Milky Way (MW) and its halo, the intergalactic medium (IGM), and the local environment of the host galaxy \cite{macquart2020census}.
The total observed dispersion measure, ($\mathrm{DM}_\mathrm{Tot}$), can therefore be expressed as
\begin{equation}  
    \mathrm{DM}_\mathrm{Tot} = \mathrm{DM}_\mathrm{MW} + \mathrm{DM}_\mathrm{MW,Halo} + \mathrm{DM}_\mathrm{IGM} + \frac{\mathrm{DM}_\mathrm{Host}}{(1+z)},
    \label{eq:DMtot}
\end{equation}

where,
$\mathrm{DM}_{\mathrm{MW}}$ refers to the MW ISM, $\mathrm{DM}_{\mathrm{MW,Halo}}$ represents the contribution from the Galactic halo, $\mathrm{DM}_{\mathrm{IGM}}$ denotes the IGM, $\mathrm{DM}_{\mathrm{Host}}$ accounts for the host galaxy or neutron star (NS) with the appropriate redshift ($z$).

 FRBs were first discovered in the archival data from the Parkes Radio Telescope in 2007 \cite{2007Sci...318..777L}. Over the past decade, the observation capabilities have dramatically increased across the world. Observatories such as the Canadian Hydrogen Intensity Mapping Experiment (CHIME) \cite{Amiri_2018} in Canada and the Five hundred meter Aperture Spherical radio Telescope (FAST) \cite{Jiang_2022} in China have detected a larger number of FRBs, some of which are repeating. In addition, high precision interferometric arrays like the Australian Square Kilometre Array Pathfinder (ASKAP) \cite{2020ApJ...895L..37B} and Deep Synoptic Array-110 (DSA-110) \cite{2024AAS...24331903S} have enabled accurate localization of FRBs, allowing secure host-galaxy identification and redshift measurements.

These observations show FRBs as a complementary probe of the cosmic baryon distribution, alongside traditional observables such as the cosmic microwave background (CMB) and quasar absorption systems \cite{PhysRevD.89.107303}. Measurements of FRB DM provide compelling evidence for the presence of missing baryons in the intergalactic medium (IGM) \cite{Yang_2022}. FRBs also offer a unique laboratory for testing fundamental physics through precision studies of radio wave propagation over cosmological distances.

Although the origin of the FRBs is still under active investigation, some studies suggest that NS, and in particular magnetars, are among the leading candidates for FRB progenitors due to their compactness, possession of extremely high energy, and ultra-strong magnetic fields \cite{refId0,Hessels_2017, particles6010025, 10.1093/mnras/staa2128}. As FRB signals propagate through the magnetosphere of the NS prior to entering the IGM, they may experience additional propagation effects. In particular, axion-photon interactions in strong magnetic fields can induce subtle, frequency dependent modifications to photon propagation, potentially imprinting observable signatures in the arrival times of radio pulses \cite{Prabhu_2023,PhysRevD.103.043015,PhysRevD.104.103030}. This motivates a detailed examination of the DM contribution arising from the magnetospheric environment, which we discuss in the following sections.

Axions and the axion like particles (ALP) are among the most well motivated candidates for physics beyond the Standard Model \cite{PhysRevD.105.023017}. Originally proposed to solve the strong CP problem in quantum chromodynamics (QCD) \cite{PhysRevLett.38.1440}, axions naturally arise in many extensions of Standard Model \cite{universe8050253} and are also considered as potential dark matter candidates \cite{Berghaus_2025}. In the presence of a strong magnetic field, axions can convert into photons via the axion-photon coupling\cite{PhysRevD.104.103030}. The interaction Lagrangian for this process is given by \cite{PhysRevD.37.1237,PhysRevD.111.063021}
\begin{equation}
\mathcal{L}_{a\gamma\gamma} = -\frac{g_{a\gamma\gamma}}{4} \, a F^{\mu\nu} \tilde{F}_{\mu\nu} = g_{a\gamma\gamma} \, a \, \mathbf{E} \cdot \mathbf{B}.
\end{equation}

For the axion, the mass and coupling are not independent parameters but are related through the Peccei-Quinn symmetry breaking scale $f_a$ \cite{Berghaus_2025}
\begin{equation} 
m_a = \sqrt{\frac{m_u m_d}{(m_u + m_d)^2}} \frac{m_\pi^2 f_\pi^2}{f_a^2} \simeq 0.0057 \, \frac{10^9~\text{GeV}}{f_a}~\text{eV},
\end{equation}
with the axion-photon coupling given by   
\begin{equation}
g_{a\gamma\gamma} = \frac{\alpha}{2\pi f_a} \left( \frac{E}{N} - 1.92 \right),
\end{equation}
where \( E/N = 8/3 \) for the Dine–Fischler–Srednicki–Zhitnitsky (DFSZ) model  \cite{Dine:1981rt,Zhitnitsky:1980tq}.

In this work, we analyze FRB DM and redshift data compiled from multiple modern radio observatories to investigate the impact of axion-photon mixing on the observed DM. By modeling the total DM as a sum of standard astrophysical contributions and a potential axion-induced term, we perform a Markov Chain Monte Carlo (MCMC) analysis to constrain axion parameters. This approach allows FRBs to serve as complementary probes of axion physics, alongside laboratory and astrophysical searches, while simultaneously contributing to our understanding of baryons in the intergalactic medium.

This paper is organized as follows. In Sec. \ref{Framework}, we discussed the theoretical framework of NS Magnetospheres. In Sec. \ref{DM}, we discuss the various contributions to the DM. In Sec. \ref{Data and methodology}, we discuss our analysis methods. In Sec. \ref{Results}, we discuss the results obtained by fitting our model. Finally, we present our conclusion in the Sec. \ref{conclusion}.

\section{Theoretical Framework} \label{Framework}
\subsection{Neutron Star Magnetospheres}

 In order to describe photon propagation in the vicinity of the FRB source, we model the magnetospheric environment of a NS using standard assumptions commonly adopted in pulsar and magnetar studies. We consider a canonical NS radius  $R_{\rm NS}\simeq10^{6}\,\mathrm{cm}$ and assume a dipolar magnetic field configuration. We assume a  surface magnetic field strength is roughly $B_{0} \sim 10^{15}\,\mathrm{G}$, consistent with highly magnetized  NS model
 \cite{Olausen_2014,Nakagawa_2009,lorimer2005handbook}.

 Matter content around the NS magnetosphere is completely linked with the rotation of the star. The entire magnetosphere's plasma number density, including axion, is described by Goldreich-Julian (GJ) model (for details see, e.g.,\cite{Goldreich1969,Witte_2021})

\begin{equation}
n_e = \frac{2\,\boldsymbol{\Omega} \cdot \mathbf{B} / e}
{1 - \Omega^2 r^2 \sin^2\theta}
\simeq \frac{2 \Omega B_z}{e},
\end{equation}

where $\boldsymbol{\Omega}$ is the stellar angular velocity $\Omega = 2\pi/P$ and $\mathbf{B}$ is the local magnetic field. Assuming a dipolar magnetic geometry, the axial component of the field at conversion radius $r_c$ can be written as (for details see, e.g.,\cite{Berghaus_2025})

\begin{equation}
B_z =\frac{B_0}{2} \left( \frac{R_{\mathrm{NS}}}{r_c} \right)^3 \left[ 3 \cos\theta \, (\hat{\mathbf{m}} \cdot \hat{\mathbf{r}}) - \cos\theta_m \right],
\end{equation}
where, 
\begin{equation}
r_c = R_{\rm NS}
\left[
    \frac{e\,\Omega\,B_0}{m_a^2\,m_e}
    \left(
        3\cos\theta\,(\hat{\mathbf{m}}\!\cdot\!\hat{\mathbf{r}})
        - \cos\theta_m
    \right)
\right]^{1/3},
\end{equation}

\begin{equation}
r_c \simeq 36~{\rm km}\,
\left(\frac{B_0}{10^{12}\,{\rm G}}\right)^{1/3}
\left(\frac{\Omega}{2\pi\,{\rm Hz}}\right)^{1/3}
\left(\frac{m_a}{\mu{\rm eV}}\right)^{-2/3} \\
\left[
3\cos\theta\,(\hat{\mathbf{m}}\!\cdot\!\hat{\mathbf{r}})
- \cos\theta_m
\right]^{1/3}.
\end{equation}

Here $B_0$ is the surface polar field strength, ${m_a}$ is mass of the axion and $\theta_m$ is the inclination between the magnetic and rotational axes, by adapting the standard practice, we assume the polar angle
and magnetic inclination are fixed at $\theta=\pi/4$ and
$\theta_m=\pi/6$, respectively \cite{PhysRevD.104.103030}.

\begin{figure*}
\centering
\includegraphics[width=0.85\linewidth]{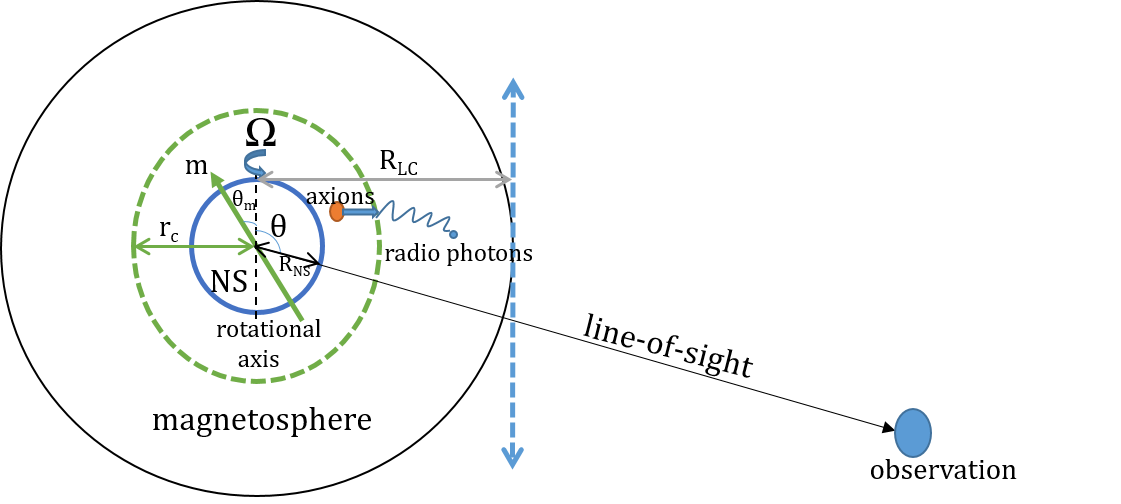}
\caption{Schematic illustration of the NS magnetosphere and the geometry relevant for axion-photon mixing. The dashed and solid circles indicate the conversion radius $r_{\rm c}$ and the light-cylinder radius $R_{\rm LC}$ respectively.  This schematic figure is inspired from \cite{Berghaus_2025}. }
\label{magnetosphere}
\end{figure*}

Figure~\ref{magnetosphere} illustrates the magnetospheric geometry adopted in this work and the characteristic scales relevant for axion-photon mixing. Axions mix with photons within the magnetosphere, with efficient conversion occurring near the  $r_c$, after which radio photons propagate outward toward the observer. The spatial extent of the closed magnetosphere is limited by the light-cylinder radius $R_{\rm LC}$, defined as the distance from the rotation axis at which the corotation velocity equals the speed of light.
This radius is given by
\begin{equation}
R_{\rm LC} = \frac{c}{\Omega} = \frac{c\,P}{2\pi},
\end{equation}
and provides a natural outer boundary for corotating plasma.
Within $R_{\rm LC}$, magnetic field lines are predominantly closed and the plasma is approximately locked into corotation, whereas beyond this radius the field structure transitions to an open configuration. The radio photons in this region can experience dispersion as a result of axion-photon mixing. To verify this, we must analyze the DM within the magnetospheric zone. This is, however, discussed in section ~\ref{DM}.

\section{Dispersion Measure} \label{DM}
The DM quantifies the integrated column density of free electrons encountered by a radio signal along its propagation path and is defined as \cite{kulkarni2020dispersionmeasureconfusionconstants}
\begin{equation}
\mathrm{DM} = \int n_e \, dl ,
\end{equation}
where $n_e$ is the electron number density and $dl$ is the differential path length.
For FRBs, the observed DM reflects contributions from multiple physically distinct regions between the source and the observer.

Accordingly, the total observed $\mathrm{DM}_\mathrm{obs}$ can be decomposed as \cite{Wang_2023}
\begin{equation}  
    \mathrm{DM}_\mathrm{obs} = \mathrm{DM}_\mathrm{MW} + \mathrm{DM}_\mathrm{MW,Halo} + \mathrm{DM}_\mathrm{IGM} + \frac{\mathrm{DM}\mathrm{Host}}{(1+z)},
\end{equation}

the factor $(1+z)^{-1}$ accounts for cosmological time dilation affecting dispersion in the host frame.

\subsection{Galactic Dispersion Measure}

The Galactic contribution to the DM originates from free electrons distributed throughout MW region.
In this work, the DM in MW is estimated using the Yao--Manchester--Wang 2016 electron density model (YMW16) \cite{Yao_2017}, which provides a three-dimensional description of the Galactic free-electron distribution constrained by pulsar observations.

For each FRB line of sight, the YMW16 model yields a direction-dependent value of $\mathrm{DM}_{\mathrm{MW}}$, incorporating contributions from the thin disk, thick disk, spiral arms, and local features.
This approach allows the Galactic foreground to be subtracted on an event-by-event basis, minimizing systematic uncertainties associated with assuming an average MW contribution.

In addition to the interstellar component, we include a contribution from the extended MW halo.
Following recent observational and theoretical studies, the halo DM is treated as a constant term with a fiducial value of 
$
\mathrm{DM}_{\mathrm{MW,halo}} = 65 \pm 15~\mathrm{pc~cm^{-3}} 
$
\cite{10.1093/mnrasl/slac022},
which is added to the YMW16 interstellar estimate.

\subsection{Host Dispersion Measure}

The DM associated with the FRB host galaxy and its immediate environment is grouped into a single term,
\begin{equation}
\mathrm{DM}_{\mathrm{host}} = \mathrm{DM}_{\mathrm{mag}} + \mathrm{DM}_{\mathrm{local}},
\end{equation}
where $\mathrm{DM}_{\mathrm{mag}}$ represents the contribution from the magnetospheric region surrounding the NS, and $\mathrm{DM}_{\mathrm{local}}$ accounts for plasma in the host galaxy and near-source environment.

The magnetospheric contribution is evaluated by integrating the electron number density along the photon trajectory within the magnetosphere,
\begin{equation}
\mathrm{DM}_{\mathrm{mag}} =
\int_{R_{\mathrm{NS}}}^{r_c} n_e^{(\mathrm{in})}(r)\,dr
+
\int_{r_c}^{R_{\mathrm{LC}}} n_e^{(\mathrm{out})}(r)\,dr ,
\end{equation}
the functions $n_e^{(\mathrm{in})}$ and $n_e^{(\mathrm{out})}$ describe the plasma density profiles in the inner and outer $r_c$ , respectively. The remaining local contribution $\mathrm{DM}_{\mathrm{local}}$ captures dispersion arising from the host galaxy ISM.
Given the diversity of FRB host environments, this term is treated as a free parameter in the statistical analysis.

\subsection{Dispersion Measure in IGM}

The IGM contribution to DM is determined by the cosmological distribution of ionized baryons.
 The mean DM contributed by the IGM at 
$z$ can be expressed as (for details see, e.g.,\cite{macquart2020census,Deng_2014,Lemos_2025})
\begin{equation}
{\rm DM}_{\rm IGM}(z)
=
\frac{3c\,\Omega_b H_0^2}{8\pi G m_p}
\int_0^z
\frac{(1+z')\,f_{\rm IGM}(z')\,\chi(z')}{H(z')}\,dz',
\end{equation}
where $c$ denotes the speed of light, $\Omega_b = 0.0486$ is the present-day
baryon density parameter \cite{macquart2020census}, $H_0 = 67.4\,\mathrm{km\,s^{-1}\,Mpc^{-1}}$ is the Hubble constant \cite{Planck2018}, $G$ is the gravitational constant, and $m_p$ is the proton mass. The function
$f_{\rm IGM}(z)$ represents the fraction of baryons residing in the
IGM, while $H(z)$ is the Hubble expansion rate at $z$.

The quantity $\chi(z)$ corresponds to the number of free electrons per
baryon and is given by
$\chi(z)=Y_{\rm H}\chi_{e,{\rm H}}(z)+Y_{\rm He}\chi_{e,{\rm He}}(z)$,
where $Y_{\rm H}=3/4$ and $Y_{\rm He}=1/4$ are the mass fractions of
hydrogen and helium, respectively. For $z\lesssim3$, both
hydrogen and helium are assumed to be fully ionized, such that
$\chi_{e,{\rm H}}(z)=\chi_{e,{\rm He}}(z)=1$.
We adopt the $\Lambda$CDM  model
throughout this work \cite{Condon_2018}. Within this framework, the expansion history of
the Universe determines the $z$ dependence of cosmological
distances and the mean baryon density, which are essential for modeling
the IGM contribution to the DM of FRBs \cite{Wang_2023}. This formulation provides the dominant contribution to the observed DM at moderate and high $z$ and enables FRBs to act as tracers of the cosmic baryon distribution.

\section{Data and Methodology} \label{Data and methodology}

To obtain robust constraints on the model parameters governing the DM of FRBs, we adopt a Bayesian inference framework based on Markov Chain Monte Carlo (MCMC) \cite{Hogg_2018}. This approach allows the joint posterior distribution of all free parameters to be explored efficiently.  We adopt 125 localized FRBs as $\mathrm{DM}_{\rm obs}$, which are listed in Appendix A, and the total DM uncertainty is modeled as 

\begin{equation}
\sigma_{\mathrm{tot}}^2
=
\sigma_{\mathrm{obs}}^2
+
\sigma_{\mathrm{MW}}^2
+
\sigma_{\mathrm{MW,Halo}}^2
+
\sigma_{\mathrm{IGM}}^2
+
\left( \frac{\sigma_{\mathrm{host}}}{1+z} \right)^2 ,
\end{equation}

where $\sigma_{\mathrm{obs}}$ denotes the measurement uncertainty, 
$\sigma_{\mathrm{MW}}$ and $\sigma_{\mathrm{MW,Halo}}$ represent the uncertainties associated with the MW disk \cite{Yao_2017} and halo contributions \cite{Wang_2023}, respectively. 
Both Galactic components are modeled with $30\%$ fractional uncertainties relative to their estimated values. 
The IGM uncertainty, $\sigma_{\mathrm{IGM}}$, is assigned a $20\%$ fractional error to account for line-of-sight fluctuations \cite{Jaroszynski2020}. For the host galaxy contribution, we adopt a fiducial uncertainty 
$\sigma_{\rm host} = 100 \, \mathrm{pc\,cm^{-3}}$, consistent with estimates from localized FRBs \cite{macquart2020census}. Since most FRB catalogs do not explicitly report measurement errors, we adopt an uncertainty randomly drawn from the Gaussian distribution by using the $\sigma_\mathrm{obs}$ of the other FRBs
\cite{Lemos_2025, 2019AARv..27....4P, Jaroszynski2020, 10.1093/mnras/stac2524}. All uncertainty components are assumed to be independent and Gaussian-distributed. We have verified that moderate variations in these assumed fractional uncertainties do not qualitatively affect our main parameter constraints.

The parametric model DM at  $z$ is given by \cite{Wang_2023}

\begin{equation}
\mathrm{DM}_{\rm model}(z) =
\mathrm{DM}_{\rm IGM}(z,f_{\rm IGM})
+ \mathrm{DM}_{\rm MW}
+ \mathrm{DM}_{\rm MW,Halo} \\
+ \frac{\mathrm{DM}_{\rm mag}(m_a,g_{a\gamma\gamma})+\mathrm{DM}_{\rm local}}{1+z}.
\label{eq. 16}
\end{equation}

The likelihood is constructed from a chi-square statistic,
\begin{equation}
\chi^2(\boldsymbol{\theta}) =
\sum_i
\frac{\left[\mathrm{DM}_{{\rm obs},i}-\mathrm{DM}_{\rm model}(z_i)\right]^2}
{\sigma_{tot,i}^2},
\end{equation}
while all other parameters are assigned uniform priors within physically motivated, which is discussed in Table \ref{tab:priors}. 
\begin{table}[ht]
\centering
\caption{Prior ranges adopted in the MCMC analysis. Logarithmic priors are uniform in $\log_{10}$.}
\begin{tabular}{lc}
\hline\hline
Parameter & Prior Range \\
\hline
$\log_{10}(m_a/\mu{\rm eV})$ 
& $[-3,\; 2]$ \\

$\log_{10}(m_a/\mu{\rm eV})$
& $[-20,\; -10]$ \\

$DM_{\rm local}$ (pc cm$^{-3}$) 
& $[0,\; 300)$ \\

$f_{\rm IGM}$ 
& $[0.70,\; 0.94]$ \\
\hline\hline
\end{tabular}
\label{tab:priors}
\end{table}

The parameter vector sampled in the analysis consists of the logarithm of the axion mass, $\log_{10}(m_a/\mu{\rm eV})$, logarithm of the axion-photon coupling, $\log_{10}(m_a/\mu{\rm eV})$, the local dispersion contribution $\mathrm{DM}_{\rm local}$, and the intergalactic baryon fraction $f_{\rm IGM}$.

 To initialize the MCMC chains, a preliminary chi-square minimization is performed using an Nelder-Mead algorithm. A small fraction of the sample exhibits DM that is inconsistent with the bulk DM-$z$ relation at the $3\sigma$ level. These events are excluded from the refitting procedure to avoid poorly constrained. The resulting fit is insensitive to the removal of individual outliers.
The resulting best-fit parameter values are used as starting points for the walkers, with small random perturbations applied to ensure adequate exploration of parameter space.
This procedure improves convergence efficiency while avoiding bias toward local minima. Using the DM model together with the full 
uncertainty, we evaluate the likelihood 
and sample the corresponding posterior distributions of the 
model parameters. The resulting constraints are summarized 
and discussed in the Sec \ref{Results}.

\begin{figure}[ht!]
\centering
\includegraphics[width=0.7\columnwidth]{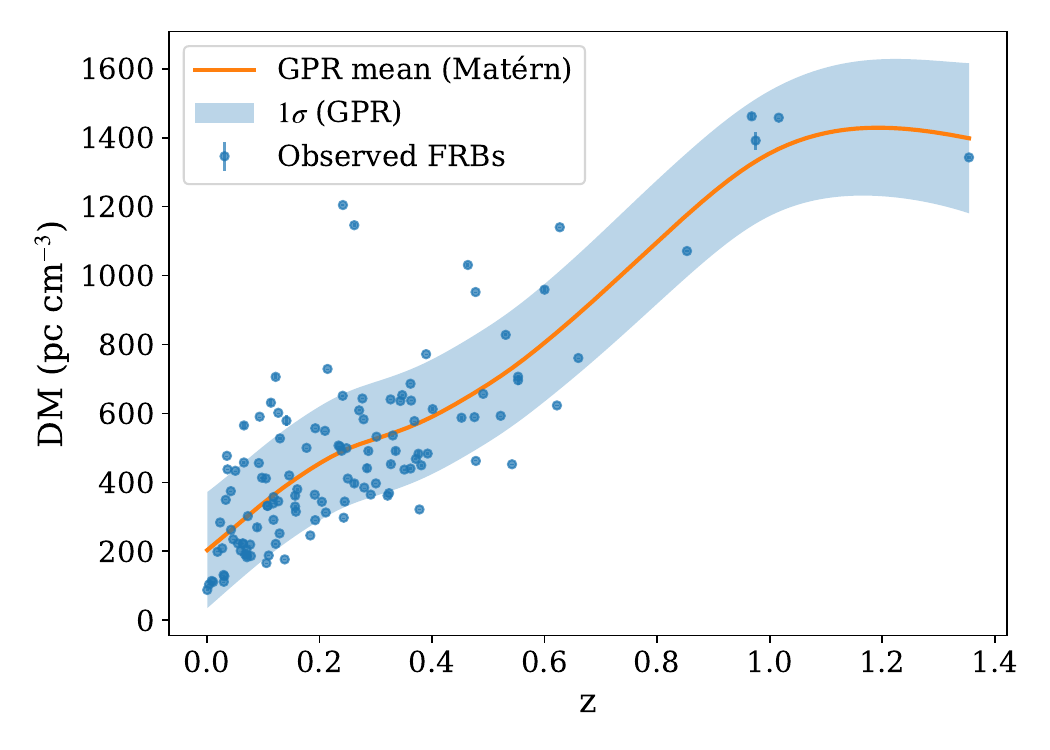}
\caption{ DM as a function of $z$ for the FRB sample.
The solid line shows the Gaussian Process posterior mean obtained with a
Matérn kernel, and the shaded region indicates the $1\sigma$ confidence
band. DM values used in the
reconstruction are listed in Appendix A.
}
\label{GPR}
\end{figure}
\subsection{Non-parametric Reconstruction as a Robustness Test}

To assess the sensitivity of our results to assumptions about the cosmological modeling of the DM, we repeat the analysis using a non-parametric Gaussian Process Regression (GPR) reconstruction of the DM-$z$ relation \cite{rasmussen2006gaussian,annurev:/content/journals/10.1146/annurev-astro-052920-103508} to the FRB dataset.
This approach does not rely on a specific parametric form for $\mathrm{DM}_{\rm IGM}(z)$ and therefore provides a complementary consistency check on the axion-photon mixing constraints derived in the previous subsection.

We emphasize that the GPR analysis is intended as a robustness test of the parametric results rather than as an independent detection framework. Since the reconstruction is trained directly on the observed FRB dataset with measured $z$ spanning 
$0 \lesssim z \lesssim 1.3$, and observed DM in the range $\sim 100$–$ 1600~\mathrm{pc\,cm^{-3}}$, any smooth $z$-dependent component, including a potential axion induced contribution, may be partially absorbed into the reconstructed mean function. The resulting constraints should therefore be interpreted as conservative bounds on deviations from the empirical DM-$z$ relation.

The regression is performed using a Matérn covariance kernel \cite{rasmussen2006gaussian,wang2023parameterizationmaterncovariancefunction}, multiplied by a constant amplitude and supplemented by a white-noise term. 
Compared to kernels that assume infinitely differentiable functions, the Matérn form allows controlled departures from perfect smoothness and is therefore well suited to FRB data, where astrophysical scatter and environmental variations introduce moderate irregularities. 
The white-noise component accounts for intrinsic dispersion and unmodeled uncertainties. 
The resulting reduced $\chi^2$ is close to unity, indicating that the reconstruction provides an adequate description of the data within the inferred uncertainties. The reconstructed DM-$z$ relation is shown in Fig.~\ref{GPR}. 

The GPR-reconstructed mean function, $\mathrm{DM}_{\rm GPR}(z)$, is incorporated into the likelihood as a baseline model against which additional physical contributions are tested. 
The total modeled DM is written as
\begin{equation}
\mathrm{DM}_{\rm model}(z) =\;
 \mathrm{DM}_{\rm GPR}(z) \\
+ \frac{\mathrm{DM}_{\rm mag}(m_a, g_{a\gamma\gamma})
+ \mathrm{DM}_{\rm residual}}{1+z}.
\end{equation}
In this formulation, the parameter previously denoted $DM_{\rm local}$ is reinterpreted as a residual contribution, $DM_{\rm residual}$. 
This quantity does not represent the full host galaxy DM, rather, it parameterizes a systematic offset relative to the GPR-inferred mean relation. 
If the dominant $z$-dependent behavior is already captured by the reconstruction, this residual term is expected to be small.

The total variance in the GPR-based analysis is modeled as
\begin{equation}
\sigma_{\mathrm{tot}}^2
=
\sigma_{\mathrm{GPR}}^2
+
\left( \frac{\sigma_{\mathrm{host}}}{1+z} \right)^2 ,
\end{equation}
where $\sigma_{\mathrm{GPR}}$ denotes the predictive uncertainty returned by the regression. 
All remaining parameter estimation is performed following the same Bayesian methodology described in the previous method. 
The resulting constraints on $m_a\ (\mu{\rm eV})$, $g_{a\gamma\gamma}\ ({\rm GeV}^{-1})$, and $DM_{\rm residual}$ are summarized in Table~\ref{tab:GPR_MCMC}.

 \section{Results and Discussion} \label{Results}

 \begin{figure*}[htbp]
    \centering   
        \centering
        \includegraphics[width=0.85\textwidth]{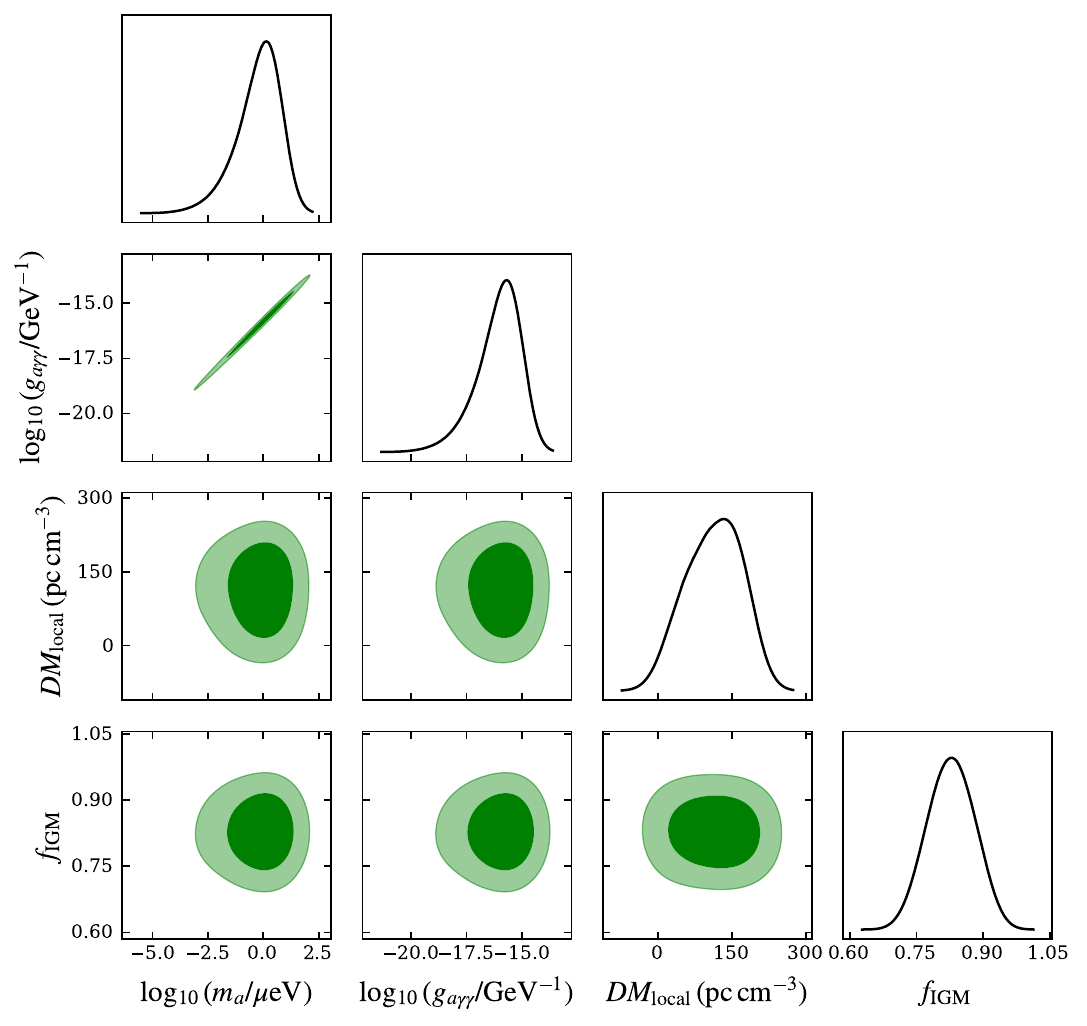}
    \caption{Marginalized posterior distributions for the model parameters 
$m_a \ (\mu{\rm eV})$, $g_{a\gamma\gamma} \ ({\rm GeV}^{-1})$, 
$\mathrm{DM}_{\rm local}$, and $f_{\rm IGM}$ inferred from the MCMC analysis of FRBs.
The 1D and 2D contours correspond to the 68\% and 95\% credible regions.}
\label{mcmc}  
\end{figure*}

Posterior constraints are summarized using the median values of each parameter and the corresponding 16th and 84th percentiles, which define the central credible interval.
Two-dimensional marginalized distributions are visualized using confidence contours enclosing 68\% and 95\% of the posterior probability, providing a clear representation of parameter correlations.
The resulting posterior distributions are shown in Fig.~\ref{mcmc}. Table~\ref{tab:mcmc_params} summarizes the marginalized posterior
constraints on the model parameters inferred from the MCMC analysis. The parameters
$m_a$ and $g_{a\gamma\gamma}$ are reasonably  constrained,
with credible intervals that are consistent with the sensitivity range
of the FRB sample. The inferred value of the IGM baryon
fraction is also tightly bounded and agrees with expectations from
cosmological studies \cite{10.1093/mnrasl/slaa070}.

\begin{table}[htbp]
\centering
\setlength{\tabcolsep}{12pt}
\renewcommand{\arraystretch}{1.5} 
\caption{Best-fit of DM model parameters and $1\sigma$ uncertainties obtained from the MCMC analysis.}
\label{tab:mcmc_params}
\begin{tabular}{lc}
\hline\hline
\textbf{Parameter} & \textbf{Value} \\
\hline
$m_a \, (\mu{\rm eV})$ 
& $1.16^{+4.40}_{-1.08}$ \\

$g_{a\gamma\gamma} \, ({\rm GeV}^{-1})$ 
& $(1.76^{+6.69}_{-1.64}) \times 10^{-16}$ \\

$DM_{\rm local} \, ({\rm pc\,cm^{-3}})$ 
& $119.6^{+58.8}_{-75.4}$ \\

$f_{\rm IGM}$ 
& $0.837^{+0.053}_{-0.056}$ \\
\hline\hline
\end{tabular}
\end{table}

In contrast, the $DM_{\rm local}$ contribution
remains comparatively weakly constrained. This
reflects the limited knowledge of the plasma properties in the 
environment of FRB sources, as well as the absence of a detailed
source-specific model for host and circumburst contributions. As a
result, the posterior distribution of $DM_{\rm local}$ exhibits a
broader uncertainty relative to the other parameters.

\begin{figure}[ht!]
\centering
\includegraphics[width=0.7\columnwidth]{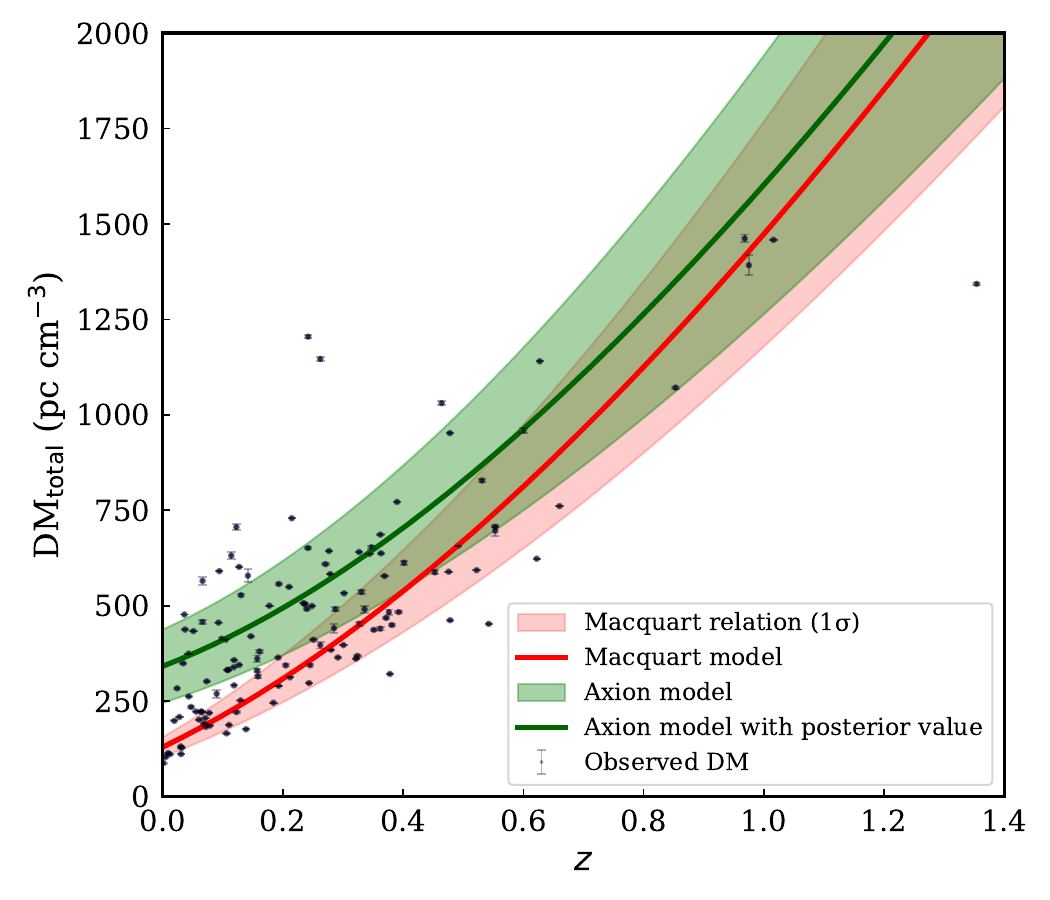}
\caption{ Observed DM of 125FRBs as a function of $z$ . The red curve shows the standard Macquart relation with a $1\sigma$ scatter band, adopting fixed foreground contributions commonly used in the literature. The green curve indicates the posterior mean prediction of the axion model obtained in this work, with the shaded region representing the corresponding $1\sigma$ uncertainty.
}
\label{Mac vs Axion}
\end{figure}

Figure~\ref{Mac vs Axion} provides a reference comparison for the observed DM. We plot the expected DM-$z$ relation based on the standard Macquart relation, shown as the red curve. In this reference model we adopt fixed contributions from the Milky Way and host galaxy, following the typical assumptions used in the literature \cite{macquart2020census}. Specifically, we take
$DM_{\rm MW,ISM}=30 \ {\rm pc \ cm^{-3}}$,
$DM_{\rm MW,halo}=50 \ {\rm pc \ cm^{-3}}$,
and a rest-frame host contribution
$DM_{\rm host}=50 \ {\rm pc \ cm^{-3}}$.
These values are commonly used to illustrate the average cosmological DM-$z$ relation and are shown here for comparison only.

In contrast, the axion model (shown as the green curve) presented in this work adopts a more data-driven treatment of the foreground contributions. The Galactic contribution $DM_{\rm MW}$ is estimated individually for each burst using the YMW16 model, while the halo contribution is taken to be $DM_{\rm MW,halo}=65\ {\rm pc \ cm^{-3}}$ \cite{10.1093/mnrasl/slac022}. The host-galaxy contribution is not fixed but is instead represented by the posterior parameter $DM_{\rm local}$, which effectively captures the combined host and local-environment contribution inferred from the data.
Most FRBs in the current sample broadly follow the Macquart DM-$z$ relation, reflecting the dominant contribution of the intergalactic medium to the observed dispersion. The comparison shown in Fig.~\ref{Mac vs Axion} illustrates that the axion model yields a DM-$z$ relation that is broadly consistent with the observed distribution of FRBs while allowing additional dispersion contributions arising from axion-photon mixing in NS magnetospheres.

\begin{figure}[ht!]
\centering
\includegraphics[width=0.7\columnwidth]{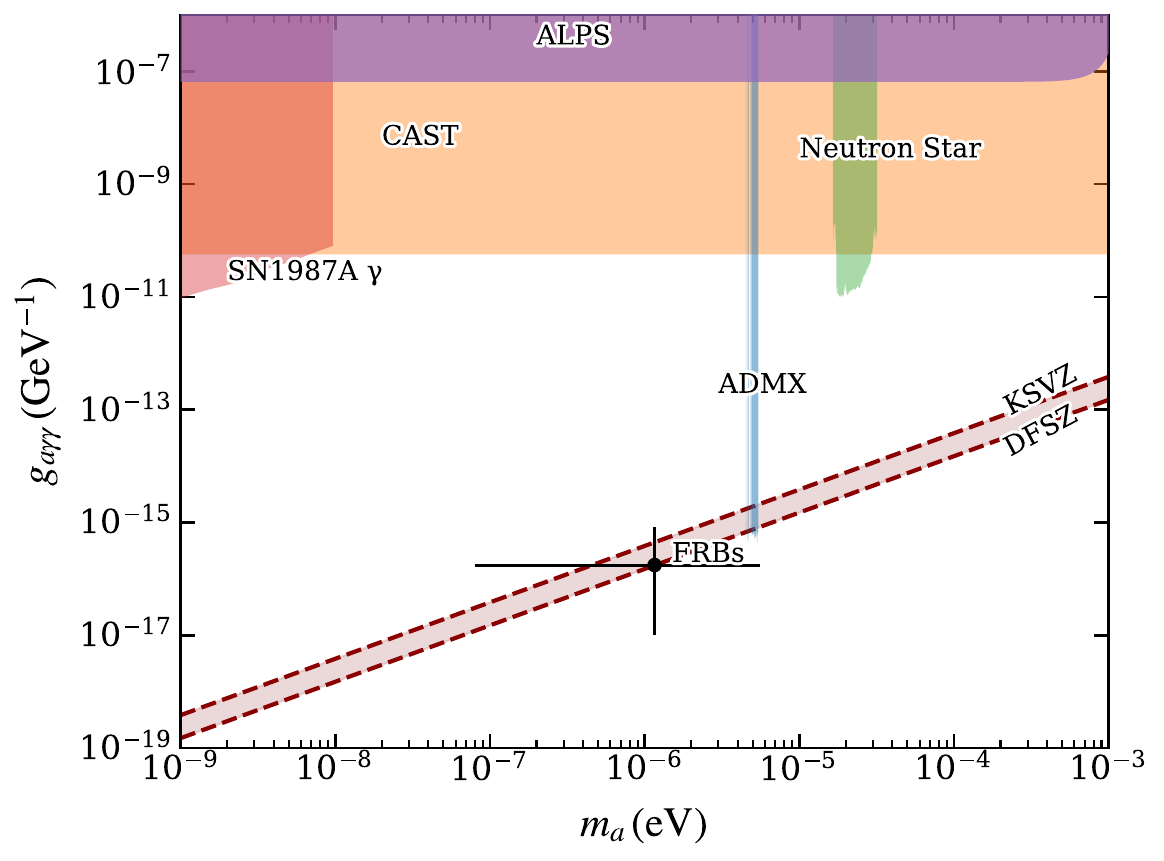}
\caption{The schematic bounds approximately shows the summary of existing constraints on the axion-photon coupling compared with the FRB-derived limits obtained in this work. All other bounds we obtain from the AxionLimits compilation \cite{AxionLimits}.}
\label{Bounds}
\end{figure}

In Figure~\ref{Bounds}, we summarizes existing constraints on the axion-photon coupling as a function of the axion mass and compares them with the limits derived in this work.
The shaded diagonal brown colour band corresponds to the theoretically motivated QCD axion region, bounded by the KSVZ and DFSZ models, which reflects the fixed relation between axion mass and coupling strength set by the Peccei--Quinn symmetry-breaking scale \cite{B_hre_2013,DINE1981199}. We include constraints from the CAST, which searches for solar axion mass and its couplings \cite{CAST:2017}.
Laboratory limits from the ADMX experiment are included \cite{ADMX:2021}, together with NS constraints derived from radio and astrophysical observations \cite{Tan:2021,2017NatPh..13..584C}.
Laboratory bound from resonant searches for dark-matter axions converting into microwave photons, sensitive in the $\mu eV$ mass range \cite{PhysRevLett.120.151301}. Bound from the non-observation of gamma rays from Supernova 1987A, limiting axion production and conversion for light axion masses \cite{Payez_2015}.
The region labeled ALPs illustrates the broader parameter space accessible to axion-like particles whose mass and coupling are not tied by the QCD relation \cite{DINE1981199}.

The point labeled “FRBs” in Fig.~\ref{Bounds} represents the parameter region inferred from the present analysis of FRB DM. This region partially overlaps with the QCD axion band, highlighting the potential sensitivity of FRB observations to very weak axion-photon couplings. The resulting FRB constraints are complementary to existing laboratory and astrophysical bounds and probe a region of parameter space that remains difficult to access with current experiments. The inferred region lies well below present experimental limits and is therefore fully consistent with existing constraints on axion-photon interactions.
\begin{figure}[ht!]
\centering
\includegraphics[width=0.7\columnwidth]{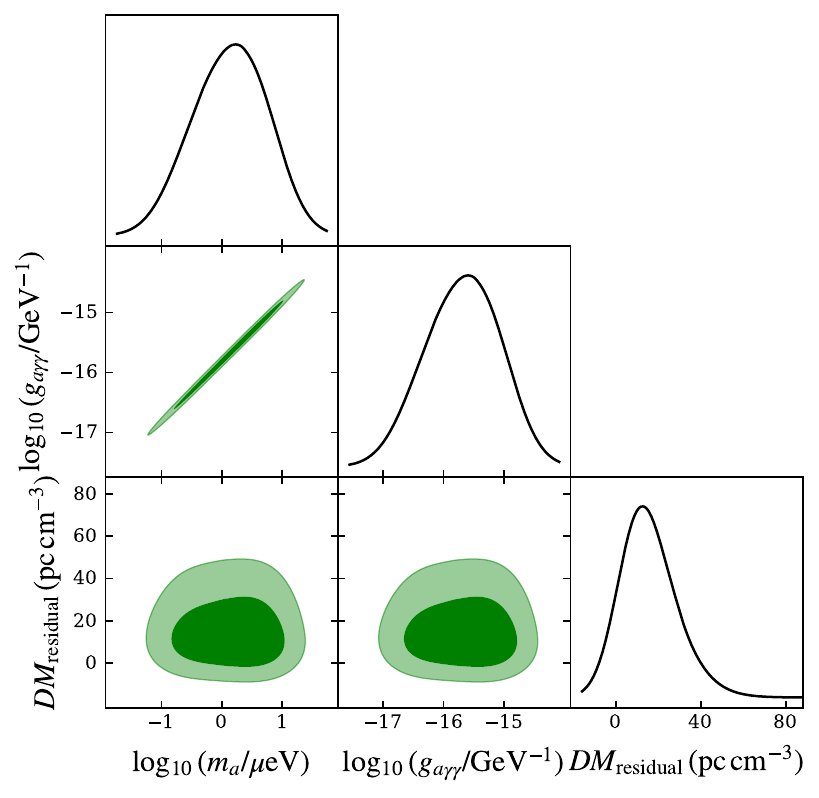}
\caption{
Marginalized posterior distributions for the  $m_a$,  $g_{a\gamma\gamma}$, and $DM_{\rm residual}$ obtained from the MCMC analysis incorporating the Gaussian Process reconstructed DM-$z$ relation. The 1D and 2D contours correspond to the 68\% and 95\% credible regions.
}
\label{GPR_MCMC}
\end{figure}

\begin{table}[t]
\centering
\setlength{\tabcolsep}{12pt}
\renewcommand{\arraystretch}{1.5} 
\caption{Best-fit of GPR reconstruction and $1\sigma$ uncertainties obtained from the MCMC analysis.}
\label{tab:GPR_MCMC}
\begin{tabular}{lc}
\hline\hline
\textbf{Parameter} & \textbf{Value} \\
\hline
$m_a$ ($\mu$eV) 
& $1.54^{+4.27}_{-1.25}$ \\

$g_{a\gamma\gamma}$ ($\mathrm{GeV}^{-1}$) 
& $(2.35^{+6.49}_{-1.90}) \times 10^{-16}$ \\

$\mathrm{DM}_{\mathrm{residual}}$ ($\mathrm{pc\,cm^{-3}}$) 
& $13.03^{+14.39}_{-9.27}$ \\
\hline
\hline
\end{tabular}
\end{table}

Figure~\ref{GPR_MCMC} shows the one dimensional marginalized posterior distributions and the corresponding two-dimensional confidence contours obtained using the GPR-based reconstruction. The numerical constraints derived from the marginalized posteriors are summarized in Table~\ref{tab:GPR_MCMC}.

We find that the axion parameters inferred within the non-parametric GPR framework remain statistically consistent with those obtained from the parametric cosmological analysis. In particular, the marginalized posteriors for $m_a$ and $g_{a\gamma\gamma}$ exhibit substantial overlap between the two approaches, indicating that the inferred bounds are not strongly sensitive to the assumed parametric form of $\mathrm{DM}_{\rm IGM}(z)$.

The notably smaller value of $DM_{\rm residual}$ relative to $DM_{\rm local}$ reflects the fact that the GPR reconstruction captures the dominant smooth redshift evolution of the dispersion measure, including the average host contribution. Consequently, only a modest systematic offset remains to be fitted in the GPR-based analysis. 
Hence, the consistency between the parametric and non-parametric results demonstrates that the derived constraints on axion mass and axion-photon coupling are stable against assumptions regarding the cosmological modeling of the dispersion measure.
Although the marginalized posterior distributions mildly disfavour $ m_a$ or $g_{a\gamma\gamma}=0$ at the 68\% confidence level in both the parametric and GPR analysis, the statistical significance does not exceed the 2$\sigma$ threshold. We therefore interpret this result as a weak preference rather than strong evidence for axion-photon mixing, and treat our findings as conservative bounds on the axion-photon mixing parameters.

\section{Conclusion} \label{conclusion}

We have investigated the effect of axion-photon mixing on the dispersion measures (DM) of fast radio bursts using a sample of 125 localized FRBs with measured redshifts. The total DM was modeled as the sum of Galactic, intergalactic medium (IGM), and source-associated contributions, with an additional magnetospheric term arising from axion-photon coupling in the neutron-star environment. A Gaussian Process reconstruction of the DM--$z$ relation was used as a data-driven consistency check on the physical model, while a Bayesian MCMC analysis was employed to infer the posterior distributions of the model parameters. 
From the parametric analysis we obtain the posterior constraints
\[
m_a = 1.16^{+4.40}_{-1.08}\,\mu{\rm eV}, \quad
g_{a\gamma\gamma} = (1.76^{+6.69}_{-1.64})\times10^{-16}\,{\rm GeV}^{-1},
\]
\[
DM_{\rm local} = 119.6^{+58.8}_{-75.4}\,{\rm pc\,cm^{-3}}, \quad
f_{\rm IGM} = 0.837^{+0.053}_{-0.056},
\]
where the quoted uncertainties correspond to the 68\% credible intervals. 
The inferred value of the IGM baryon fraction is consistent with expectations from standard cosmology \cite{10.1093/mnrasl/slaa070}, indicating that the overall dispersion budget remains compatible with the conventional cosmological interpretation.

An important source of uncertainty in the present analysis arises from the local DM contribution, which contains plasma effects within the host galaxy and the immediate source environment. In the absence of a physically motivated and source specific model for this component, $DM_{\rm local}$ remains poorly constrained and introduces a significant degeneracy with the axion-related parameters. In particular, both the magnetospheric dispersion term $DM_{\rm mag}$ and the host contribution $DM_{\rm local}$ enter the observed dispersion as $(1+z)^{-1}$, leading to a partial degeneracy between these two components. As a result, the inclusion of axion-photon mixing effectively redistributes part of the dispersion previously attributed to the host contribution into the magnetospheric term.

The FRB-derived parameter region partially overlaps with the QCD axion band, highlighting the potential of FRB observations as probes of extremely weak axion-photon couplings. These bounds are complementary to existing laboratory and astrophysical constraints and remain fully consistent with current limits. Table~\ref{tab:GPR_MCMC} summarizes the posterior constraints obtained when the analysis is repeated using the GPR-reconstructed mean DM-$z$ relation. The inferred values of the axion mass and axion-photon coupling remain statistically consistent with those obtained in the parametric analysis (Table~\ref{tab:mcmc_params}). This agreement indicates that the resulting constraints are not strongly driven by the assumed parametric form of the cosmological DM-$z$ relation.

While the marginalized posterior distributions show a mild preference for non-zero axion-photon coupling, the statistical significance remains below the threshold required to claim robust evidence. We therefore interpret the present results as providing conservative bounds rather than a detection of axion-photon mixing. Future FRB samples with improved statistics and reduced systematic uncertainties are expected to further strengthen these constraints and clarify the role of FRBs as probes of axion physics. In particular, progress will require improved modeling of the host and local contributions to FRB dispersion measures, supported by larger samples of well-localized bursts and independent constraints on their environments. Such developments will be essential for transforming FRBs from qualitative probes into quantitatively robust tools for testing axion physics and other scenarios beyond the Standard Model.
\section*{Acknowledgements}
The authors acknowledge Vellore Institute of Technology for the financial support through its Seed Grant (No.SG20230035), year 2023.

\appendix 
\label{FRB Data}
\centering
\section{Appendix: FRB Sample Used in This Work}

\setlength{\tabcolsep}{3pt}
\renewcommand{\arraystretch}{1.2}

\begin{longtable}{lccccc}
\caption{The 125 localized FRBs used in this work were compiled and updated based on the sample of  \cite{feng2025constrainingcomparingdynamicaldark}.}
\label{Tab:FRBs} \\

\hline
FRB & $z$ & $\mathrm{DM_{obs}}$ (${\rm pc\,cm^{-3}}$) 
& $\sigma_{\mathrm{obs}}$(${\rm pc\,cm^{-3}}$)
& $\mathrm{DM_{MW}^{YMW16}}$ (${\rm pc\,cm^{-3}}$) & Ref. \\
\hline
\endfirsthead

\hline
FRB & $z$ & $\mathrm{DM_{obs}}$ (${\rm pc\,cm^{-3}}$) 
& $\sigma_{\mathrm{obs}}$ (${\rm pc\,cm^{-3}}$)
& $\mathrm{DM_{MW}^{YMW16}}$ (${\rm pc\,cm^{-3}}$) & Ref. \\
\hline
\endhead

\multicolumn{5}{r}{\textit{Continued on next page}} \\
\endfoot

\endlastfoot

FRB20121102A & 0.19273 & 557.0 & 2 & 287.1 & \cite{chatterjee2017direct,Tendulkar_2017} \\
FRB20171020A & 0.00867 & 114.1 & 0.2 & 24.7 & \cite{Sherman_2024} \\
FRB20180301A & 0.33040 & 536.0 & 5 & 254.0 & \cite{Bhandari_2022} \\
FRB20180814A & 0.06800 & 190.9 & 2.52 & 107.9 & \cite{Michilli_2023} \\
FRB20180916B & 0.03370 & 349.3 & 1.62 & 324.9 & \cite{marcote2020repeating} \\
FRB20180924B & 0.32120 & 361.4 & 0.06 & 27.6 & \cite{Gordon_2023,doi:10.1126/science.aaw5903,macquart2020census,Bhandari_2020,Shannon_2025} \\
FRB20181030A & 0.00385 & 103.4 & 1.62 & 33.0 & \cite{Bhardwaj_2021} \\
FRB20181112A & 0.47550 & 589.3 & 0.03 & 29.0 & \cite{doi:10.1126/science.aay0073,macquart2020census,Bhandari_2020} \\
FRB20181220A & 0.02746 & 208.7 & 0.02 & 115.3 & \cite{Bhardwaj_2024} \\
FRB20181223C & 0.03024 & 111.6 & 0.21 & 19.1 & \cite{Bhardwaj_2024} \\
FRB20190102C & 0.29120 & 364.5 & 0.3 & 43.3 & \cite{macquart2020census,Bhandari_2020} \\
FRB20190110C & 0.12244 & 221.6 & 1.98 & 29.9 & \cite{Ibik_2024} \\
FRB20190303A & 0.06400 & 223.2 & 0.42 & 21.8 & \cite{Michilli_2023} \\
FRB20190418A & 0.07132 & 182.8 & 1.42 & 85.8 & \cite{Bhardwaj_2024} \\
FRB20190425A & 0.03122 & 127.8 & 0.34 & 38.7 & \cite{Bhardwaj_2024} \\
FRB20190520B & 0.24180 & 1204.7 & 4 & 50.2 & \cite{Gordon_2023,niu2022repeating} \\
FRB20190523A & 0.66000 & 760.8 & 0.6 & 29.9 & \cite{ravi2019fast} \\
FRB20190608B & 0.11778 & 338.7 & 0.5 & 26.6 & \cite{Gordon_2023,macquart2020census,Bhandari_2020,Shannon_2025} \\
FRB20190611B & 0.37780 & 321.4 & 0.2 & 43.7 & \cite{Heintz_2020} \\
FRB20190614D & 0.60000 & 959.2 & 7.43 & 108.7 & \cite{Law_2020} \\
FRB20190711A & 0.52200 & 593.1 & 0.4 & 42.6 & \cite{Heintz_2020} \\
FRB20190714A & 0.23650 & 504.1 & 2 & 31.2 & \cite{Heintz_2020,Shannon_2025} \\
FRB20191001A & 0.23400 & 506.9 & 0.04 & 31.1 & \cite{Heintz_2020,Shannon_2025} \\
FRB20191106C & 0.10775 & 332.2 & 0.03 & 20.5 & \cite{Ibik_2024} \\
FRB20191228A & 0.24320 & 297.5 & 0.05 & 20.1 & \cite{Bhandari_2020,Shannon_2025} \\
FRB20200223B & 0.06024 & 201.8 & 0.02 & 37.0 & \cite{Ibik_2024} \\
FRB20200430A & 0.16080 & 380.1 & 4 & 26.1 & \cite{Heintz_2020,Shannon_2025} \\
FRB20200906A & 0.36880 & 577.8 & 0.2 & 37.9 & \cite{Bhandari_2022,Shannon_2025} \\
FRB20201020E & 0.00080 & 87.8 & 1.62 & 32.2 & \cite{kirsten2022repeating} \\
FRB20201123A & 0.05070 & 433.6 & 0.04 & 162.7 & \cite{10.1093/mnras/stac1450} \\
FRB20201124A & 0.09800 & 413.5 & 3.23 & 196.6 & \cite{Fong_2021} \\
FRB20210117A & 0.21450 & 729.1 & 0.36 & 23.1 & \cite{Gordon_2023,Shannon_2025} \\
FRB20210320C & 0.27970 & 384.8 & 0.03 & 30.4 & \cite{Gordon_2023,Shannon_2025} \\
FRB20210405I & 0.06600 & 565.2 & 10.17 & 348.7 & \cite{10.1093/mnras/stad3329} \\
FRB20210410D & 0.14150 & 578.8 & 16.32 & 42.2 & \cite{Gordon_2023,10.1093/mnras/stad1839} \\
FRB20210603A & 0.17720 & 500.1 & 0.05 & 30.8 & \cite{Cassanelli_2024} \\
FRB20210807D & 0.12930 & 251.9 & 0.30 & 93.7 & \cite{Gordon_2023,Shannon_2025} \\
FRB20211127I & 0.04690 & 234.8 & 1.94 & 31.5 & \cite{Gordon_2023,Glowacki_2023,Shannon_2025} \\
FRB20211203C & 0.34390 & 636.2 & 0.06 & 48.4 & \cite{Gordon_2023,Shannon_2025} \\
FRB20211212A & 0.07070 & 206.0 & 1.14 & 27.5 & \cite{Gordon_2023,Shannon_2025} \\
FRB20220105A & 0.27850 & 583.0 & 0.05 & 20.6 & \cite{Gordon_2023,Shannon_2025} \\
FRB20220204A & 0.40120 & 612.6 & 5.04 & 46.0 & \cite{Sherman_2024,sharma2024preferential,connor2025gasrichcosmicwebrevealed} \\
FRB20220207C & 0.04304 & 262.4 & 1.70 & 83.3 & \cite{Sherman_2024,connor2025gasrichcosmicwebrevealed} \\
FRB20220208A & 0.35100 & 437.0 & 2.99 & 107.9 & \cite{Sherman_2024,sharma2024preferential,connor2025gasrichcosmicwebrevealed} \\
FRB20220222C & 0.85300 & 1071.2 & 2.86 & 42.3 & \cite{Pastor_Marazuela_2025} \\
FRB20220224C & 0.62710 & 1140.2 & 0.11 & 51.5 & \cite{Pastor_Marazuela_2025} \\
FRB20220307B & 0.24812 & 499.3 & 1.09 & 186.9 & \cite{Sherman_2024,connor2025gasrichcosmicwebrevealed} \\
FRB20220310F & 0.47796 & 462.2 & 1.36 & 39.5 & \cite{Sherman_2024,connor2025gasrichcosmicwebrevealed} \\
FRB20220319D & 0.01123 & 111.0 & 1.79 & 211.0 & \cite{Sherman_2024} \\
FRB20220330D & 0.37140 & 468.1 & 2.03 & 30.6 & \cite{Sherman_2024,sharma2024preferential,connor2025gasrichcosmicwebrevealed} \\
FRB20220418A & 0.62200 & 623.2 & 0.06 & 29.5 & \cite{Sherman_2024,connor2025gasrichcosmicwebrevealed} \\
FRB20220501C & 0.38100 & 449.5 & 3.56 & 14.0 & \cite{Shannon_2025,sharma2024preferential} \\
FRB20220506D & 0.30039 & 397.0 & 0.03 & 97.7 & \cite{Sherman_2024,sharma2024preferential,connor2025gasrichcosmicwebrevealed} \\
FRB20220509G & 0.08940 & 269.5 & 10 & 52.1 & \cite{Bhardwaj_2024,Sherman_2024,connor2025gasrichcosmicwebrevealed} \\
FRB20220529A & 0.18390 & 246.0 & 0.02 & 30.9 & \cite{Gao_2025} \\
FRB20220610A & 1.01600 & 1458.2 & 0.2 & 13.6 & \cite{Shannon_2025} \\
FRB20220717A & 0.36295 & 637.3 & 1.05 & 83.2 & \cite{10.1093/mnras/stae1652} \\
FRB20220725A & 0.19260 & 290.4 & 0.02 & 11.6 & \cite{Shannon_2025} \\
FRB20220726A & 0.36190 & 686.2 & 1.24 & 111.4 & \cite{Sherman_2024,sharma2024preferential,connor2025gasrichcosmicwebrevealed} \\
FRB20220825A & 0.24140 & 651.2 & 3.04 & 86.9 & \cite{Sherman_2024,connor2025gasrichcosmicwebrevealed} \\
FRB20220831A & 0.26200 & 1146.2 & 5.28 & 182.3 & \cite{connor2025gasrichcosmicwebrevealed} \\
FRB20220912A & 0.07710 & 219.5 & 0.02 & 122.2 & \cite{Ravi_2023} \\
FRB20220914A & 0.11390 & 631.3 & 10 & 51.1 & \cite{Sherman_2024,connor2025gasrichcosmicwebrevealed} \\
FRB20220918A & 0.49100 & 656.8 & 0.06 & 28.9 & \cite{Shannon_2025} \\
FRB20220920A & 0.15824 & 315.0 & 3.07 & 33.4 & \cite{Sherman_2024,connor2025gasrichcosmicwebrevealed} \\
FRB20221012A & 0.28467 & 441.1 & 11.99 & 50.5 & \cite{Sherman_2024,connor2025gasrichcosmicwebrevealed} \\
FRB20221027A & 0.54220 & 452.5 & 0.04 & 41.1 & \cite{Sherman_2024,sharma2024preferential,connor2025gasrichcosmicwebrevealed} \\
FRB20221029A & 0.97500 & 1391.8 & 26.01 & 36.4 & \cite{Sherman_2024,sharma2024preferential,connor2025gasrichcosmicwebrevealed} \\
FRB20221101B & 0.23950 & 491.6 & 2.55 & 192.4 & \cite{Sherman_2024,sharma2024preferential,connor2025gasrichcosmicwebrevealed} \\
FRB20221106A & 0.20440 & 343.8 & 2.87 & 31.8 & \cite{Sherman_2024,Shannon_2025} \\
FRB20221113A & 0.25050 & 411.0 & 0.89 & 115.4 & \cite{Sherman_2024,sharma2024preferential,connor2025gasrichcosmicwebrevealed} \\
FRB20221116A & 0.27640 & 643.4 & 0.06 & 196.2 & \cite{sharma2024preferential,connor2025gasrichcosmicwebrevealed} \\
FRB20221219A & 0.55300 & 706.7 & 0.07 & 38.6 & \cite{Sherman_2024,sharma2024preferential,connor2025gasrichcosmicwebrevealed} \\
FRB20230124A & 0.09390 & 590.6 & 0.05 & 31.8 & \cite{Sherman_2024,sharma2024preferential,connor2025gasrichcosmicwebrevealed} \\
FRB20230125D & 0.32650 & 640.8 & 0.06 & 190.8 & \cite{Pastor_Marazuela_2025} \\
FRB20230203A & 0.14640 & 420.1 & 1.41 & 22.9 & \cite{Amiri_2025} \\
FRB20230216A & 0.53100 & 828.0 & 4.41 & 28.1 & \cite{sharma2024preferential,connor2025gasrichcosmicwebrevealed} \\
FRB20230222A & 0.12230 & 706.1 & 7.22 & 188.1 & \cite{Amiri_2025} \\
FRB20230222B & 0.11000 & 187.8 & 0.85 & 26.3 & \cite{Amiri_2025} \\
FRB20230307A & 0.27060 & 608.9 & 2.82 & 29.5 & \cite{sharma2024preferential,connor2025gasrichcosmicwebrevealed} \\
FRB20230311A & 0.19180 & 364.3 & 2.31 & 115.7 & \cite{Sherman_2024,sharma2024preferential,connor2025gasrichcosmicwebrevealed} \\
FRB20230501A & 0.30150 & 532.5 & 0.05 & 180.2 & \cite{Sherman_2024,connor2025gasrichcosmicwebrevealed} \\
FRB20230506C & 0.38960 & 772.0 & 0.07 & 64.1 & \cite{Anna_Thomas_2025} \\
FRB20230521B & 1.35400 & 1342.9 & 3.75 & 209.7 & \cite{Shannon_2025,connor2025gasrichcosmicwebrevealed} \\
FRB20230526A & 0.15700 & 361.4 & 8.37 & 21.9 & \cite{Shannon_2025} \\
FRB20230613A & 0.39230 & 483.5 & 3.18 & 17.2 & \cite{Pastor_Marazuela_2025} \\
FRB20230626A & 0.32700 & 452.7 & 5.79 & 32.5 & \cite{Sherman_2024,sharma2024preferential,connor2025gasrichcosmicwebrevealed} \\
FRB20230628A & 0.12700 & 345.0 & 1.22 & 30.8 & \cite{Sherman_2024,sharma2024preferential,connor2025gasrichcosmicwebrevealed} \\
FRB20230703A & 0.11840 & 291.3 & 0.02 & 20.7 & \cite{Amiri_2025} \\
FRB20230708A & 0.10500 & 411.5 & 0.04 & 44.0 & \cite{Shannon_2025} \\
FRB20230712A & 0.45250 & 587.6 & 5.45 & 30.9 & \cite{Sherman_2024,sharma2024preferential,connor2025gasrichcosmicwebrevealed} \\
FRB20230718A & 0.03570 & 477.0 & 0.047 & 450.0 & \cite{Shannon_2025} \\
FRB20230730A & 0.21150 & 312.5 & 0.20 & 97.4 & \cite{Amiri_2025} \\
FRB20230808F & 0.34720 & 653.2 & 3.88 & 26.5 & \cite{Anna_Thomas_2025} \\
FRB20230814B & 0.55300 & 696.4 & 14.47 & 137.8 & \cite{connor2025gasrichcosmicwebrevealed} \\
FRB20230902A & 0.36190 & 440.1 & 4.95 & 25.5 & \cite{Shannon_2025} \\
FRB20230907D & 0.46380 & 1030.8 & 4.51 & 28.7 & \cite{Pastor_Marazuela_2025} \\
FRB20230926A & 0.05530 & 222.8 & 1.30 & 43.7 & \cite{Amiri_2025} \\
FRB20230930A & 0.09250 & 456.0 & 0.04 & 61.7 & \cite{Anna_Thomas_2025} \\
FRB20231005A & 0.07130 & 189.4 & 0.01 & 28.8 & \cite{Amiri_2025} \\
FRB20231011A & 0.07830 & 186.3 & 1.25 & 65.7 & \cite{Amiri_2025} \\
FRB20231017A & 0.24500 & 344.2 & 3.18 & 55.6 & \cite{Amiri_2025} \\
FRB20231020B & 0.47750 & 952.2 & 0.09 & 27.6 & \cite{Pastor_Marazuela_2025} \\
FRB20231025B & 0.32380 & 368.7 & 2.27 & 43.4 & \cite{Amiri_2025} \\
FRB20231120A & 0.03680 & 437.7 & 2.14 & 36.2 & \cite{Sherman_2024,sharma2024preferential,connor2025gasrichcosmicwebrevealed} \\
FRB20231123A & 0.07290 & 302.1 & 1.03 & 136.9 & \cite{Amiri_2025} \\
FRB20231123B & 0.26210 & 396.9 & 8.98 & 33.8 & \cite{Sherman_2024,sharma2024preferential,connor2025gasrichcosmicwebrevealed} \\
FRB20231128A & 0.10790 & 331.6 & 2.50 & 20.5 & \cite{Amiri_2025} \\
FRB20231204A & 0.06440 & 221.0 & 0.02 & 21.8 & \cite{Amiri_2025} \\
FRB20231206A & 0.06590 & 457.7 & 4.83 & 59.3 & \cite{Amiri_2025} \\
FRB20231220A & 0.33550 & 491.2 & 9.11 & 44.5 & \cite{connor2025gasrichcosmicwebrevealed} \\
FRB20231223C & 0.10590 & 165.8 & 0.83 & 38.6 & \cite{Amiri_2025} \\
FRB20231226A & 0.15690 & 329.9 & 4.54 & 26.7 & \cite{Shannon_2025} \\
FRB20231229A & 0.01900 & 198.5 & 1.46 & 51.8 & \cite{Amiri_2025} \\
FRB20231230A & 0.02980 & 131.4 & 0.16 & 83.3 & \cite{Amiri_2025} \\
FRB20240114A & 0.13000 & 527.6 & 3.30 & 38.8 & \cite{10.1093/mnras/stae2013} \\
FRB20240119A & 0.37600 & 483.1 & 7.09 & 31.0 & \cite{connor2025gasrichcosmicwebrevealed} \\
FRB20240123A & 0.96800 & 1462.0 & 10.39 & 113.0 & \cite{connor2025gasrichcosmicwebrevealed} \\
FRB20240201A & 0.04273 & 374.5 & 0.03 & 29.1 & \cite{Shannon_2025} \\
FRB20240209A & 0.13840 & 176.5 & 0.38 & 52.9 & \cite{Eftekhari_2025} \\
FRB20240210A & 0.02369 & 283.7 & 2.28 & 17.9 & \cite{Shannon_2025} \\
FRB20240213A & 0.11850 & 357.4 & 0.03 & 32.1 & \cite{connor2025gasrichcosmicwebrevealed} \\
FRB20240215A & 0.21000 & 549.5 & 0.05 & 42.8 & \cite{connor2025gasrichcosmicwebrevealed} \\
FRB20240229A & 0.28700 & 491.1 & 4.93 & 29.5 & \cite{connor2025gasrichcosmicwebrevealed} \\
FRB20240310A & 0.12700 & 601.8 & 0.06 & 19.8 & \cite{Shannon_2025} \\
\hline
\end{longtable}

\bibliographystyle{JHEP}
\bibliography{FRB_Ref}

\end{document}